# Special features of Wigner times delay in slow elastic electron scattering by shallow potential well

**M. Ya. Amusia[1,2] and A. S. Baltenkov[3]**

[1] *Racah Institute of Physics, the Hebrew University, Jerusalem, 91904 Israel*
[2] *Ioffe Physical-Technical Institute, St. Petersburg, 194021 Russia*
[3] *Arifov Institute of Ion-Plasma and Laser Technologies, Tashkent, 100125 Uzbekistan*

**Abstract**

We investigate specific features in the Wigner time behavior for slow electron elastic scattering by shallow potential wells. We considered two types of potentials wells, the small changes in the parameters of which lead to arising bound states in the well. It appeared that the time delay for attractive potential wells with no bound levels always has a positive value for small electron energies and changes sign after level arising in the well. At the moment of arising the times delay has a jump. The value of this jump is as more as less is the difference in the potential well depth from its critical value. The values of times delay strongly depend on geometrical sizes of potential wells.

## 1. Introduction

Eisenbud, Wigner and Smith (EWS) were the first who interpreted the energy derivative of the scattering phase shift as the time of transmission of a particle wave packet via the scattering potential [1-3]. They introduced the EWS-time delay as a quantum dynamical observable for particle resonance scattering. At the present days the experimental investigations of these times become possible. The attosecond laser pulses, in spite of some limitations due to uncertainty principle, have opened up the possibility of experimental investigation of time delay in photoionization processes and now this is a rapidly developing field of research, as is seen, for example, in [4-11] and references therein.

The paper [12] is dedicated to studying of the partial EWS-times delay in slow elastic e-$C_{60}$ scattering within the framework of a Dirac bubble potential model for the fullerene shell. We found there that the sign of partial times delay depends on the presence of the discrete level with the correspondent orbital moment $l$ in the $C_{60}$ potential well. The $l^{\text{th}}$ time delay is positive when there are no discrete $l$-levels in the well and it is negative when these levels exist, i.e. when capturing and retaining for some time of an incident particle by potential well is replaced by its pushing out at the moment of appearing in the well of a bound state.

We investigate here this specific feature in the Wigner (or EWS) times behavior by calculating the scattering phases and times delay for slow electron elastic scattering by shallow potential wells, small changes in the parameters of which lead to arising bound states in the well. We consider two types of shallow spherical rectangular potentials. The first of them is a spherical potential that is non-equal to zero inside the sphere with radius $R_o$. The second one is a potential of the spherical layer that is non-equal to zero in the space between two concentric spheres with inner $R_{in}$ and outer $R_o$ radiuses. We will call the latter spherical rectangular potential of the fullerene shell, because this model potential is widely used to describe a potential well of the $C_{60}$ fullerene shell (see [13] and references therein).

The structure of the paper is as follows. In Sections 2 and 3 the general formulas for the phase shifts by these potential wells are derived. Sections 2.1 and 3.1 describes the general equations for potential wells parameters, for which the first *s*- and *p*-discrete levels appear. Section 4 presents results of numerical calculations of phase shifts and corresponding Wigner times delay. For the first potential well the results are in Section 4.1 and for the second potential well they are in Section 4.2. Section 5 is Conclusions and Discussions.

## 2. Spherical rectangular potential well

Let the electron with the wave vector *k* in continuum move above the potential well

$$U(r)=\begin{cases}-U_0, & r<R_o,\\ 0, & r>R_o.\end{cases} \tag{1}$$

Then the electron wave vectors inside the potential well *q* and outside it *k* are connected be the following equation[1]

$$\frac{k^2}{2}+U_0=\frac{q^2}{2} \tag{2}$$

Inside the potential well, at $r<R_o$, the solution of the wave equation is a spherical Bessel function

$$P_{kl}(r)=Aqrj_l(qr), \quad . \tag{3}$$

Beyond the potential well, at $r>R_o$, the solution of the wave equation is a linear combination of the regular and irregular spherical Bessel functions [14]

$$P_{kl}(r)=Bkr[j_l(kr)\cos\delta_l-n_l(kr)\sin\delta_l], \tag{4}$$

with the following asymptotic behavior

$$j_l\big|_{x\to\infty}(x)\to\sin(x-\pi l/2)/x \text{ and } n_l\big|_{x\to\infty}(x)\to-\cos(x-\pi l/2)/x. \tag{5}$$

Matching the logarithmic derivatives of the wave functions (3) and (4) at $r=R_0$ leads to the following expression for tangent of the phase shift of electron scattering by the potential well under consideration

$$\tan\delta_l(k)=\frac{[qj_l(kR_o)-kj_l(qR_o)]j_l'(qR_o)}{qn_l(kR_o)j_l'(qR_o)-kj_l(qR_o)n_l'(qR_o)}. \tag{6}$$

[1] Throughout this paper, we use the atomic units.

Here $j_l'(\rho) \equiv dj_l(x)/dx\big|_{x=\rho}$ and $n_l'(\rho) \equiv dn_l(x)/dx\big|_{x=\rho}$. The phase shift $\delta_l(k)$ defined by Eq. (6) is a multivalued function of $k$ with the principal values being within the interval $-\pi/2 \le \delta_l(k) \le \pi/2$. The phases $\delta_l(k)$ are defined within accuracy up to the integer number of $\pi$; so they can be always made positive.

If the phase shifts are normalized by the condition $\delta_l(k \to \infty) = 0$ then according to Levinson's theorem [15] there exist the following equality

$$\delta_l(0) - \delta_l(\infty) = n\pi . \tag{7}$$

Here $n$ is the number of discrete levels with the orbital moment $l$ in the attractive field $U(r)$. Paying attention to the fact that the following formulas $j_l(x) \propto x^l$ and $n_l(x) \propto x^{-l-1}$ describes the Bessel functions at $k \to 0$, we obtain from (6) that $\delta_l(k)_{k\to 0} \propto k^{2l+1}$, as it should be in accordance with the Wigner threshold law [16].

**2.1. The first zero energy *s*- and *p*-levels in the potential well Eq. (1)**

The following formula connects the electron wave vectors inside, $q$, and outside, $k$, the well (1) for a discrete state with the binding energy $I = -\kappa^2/2$

$$U - \frac{\kappa^2}{2} = \frac{q^2}{2} . \tag{8}$$

The solutions of the Schrödinger equation that for large values of $r$ ($r>>R_o$), decrease as $e^{-r}$ are spherical Hankel functions [14]

$$P_{\kappa l}(r) = Bh_l^{(1)}(i\kappa r), \quad . \tag{9}$$

The following formulas describe explicitly the $s$- and $p$-wave functions $P_{\kappa l}(r)$

$$P_{\kappa 0}(r) = Be^{-\kappa r}, \tag{10}$$

$$P_{\kappa 1}(r) = B\left(1 + \frac{1}{\kappa r}\right)e^{-\kappa r} . \tag{11}$$

Matching the logarithmic derivatives of the wave functions (3) and (10) at the point $R_o$ results in the following equations for the potential well parameters when a discrete $s$-level with zero binding energy $\kappa^2/2 = 0$ appears in the well

$$\cot qR_o = 0 . \tag{12}$$

From (12) we obtain the following expression connecting the depth of the potential well $U_s$ with its radius when the $s$-level appears in it

$$U_{0,s} = \pi^2 / 8R_o^2 . \tag{13}$$

Performing similar calculations for the $p$-level and using the wave functions (3) and (11), we obtain the following equation for the parameters of the potential well when the $p$-level with zero binding energy appears in it

$$j_0(qR_o) = 0 . \tag{14}$$

From here for the well parameters we have the following formula

$$U_{0,p} = \pi^2 / 2R_o^2 . \tag{15}$$

**3. Spherical rectangular potential of $C_{60}$ fullerene shell**

Here we consider the next potential well. Let an electron with the wave vector $k$ move in the continuum above the potential well

$$U(r) = \begin{cases} 0, & 0 < r < R_{in}, \\ -U_0, & R_{in} < r < R_o, \\ 0, & r > R_o. \end{cases} \tag{16}$$

We will assume that the electron wave vectors inside the potential well, $q$. and outside the well, $k$, are connected by the relation (2). Matching of the logarithmic derivatives of the wave functions $P_{kl}(r)$ collected in Table 1 (the first line) at the points $R_{in}$ and $R_o$ results in the following expression for tangent of the phases of electron scattering by the fullerene shell

$$\tan\delta_l = \frac{q[j_l'(qR_o) + D_l n_l'(qR_o)]j_l(kR_o) - k[j_l(qR_o) + D_l n_l(qR_o)]j_l'(kR_o)}{q[j_l'(qR_o) + D_l n_l'(qR_o)]n_l(kR_o) - k[j_l(qR_o) + D_l n_l(qR_o)]n_l'(kR_o)}, \tag{17}$$

where the function $D_l = D_l(R_{in})$ has the form

$$D_l = -\frac{qj_l(kR_{in})j_l'(qR_{in}) - kj_l'(kR_{in})j_l(qR_{in})}{qj_l(kR_{in})n_l'(qR_{in}) - kj_l'(kR_{in})n_l(qR_{in})}. \tag{18}$$

In the limit $R_{in} \to 0$ the function $D_l$ in equation (18) goes to zero and the phase of electron scattering by the fullerene shell (17) coincides with the scattering phase (6).

**3.1. The first zero energy *s*- and *p*-levels in the potential well Eq. (15)**

The formula (8) connects, as before, the electron wave vectors inside and outside the well ($q$ and $\kappa$, respectively) with the discrete states binding energy $I = -\kappa^2/2$. The second and third lines in Table 1 give solutions of the wave equation within the three ranges under consideration. In the region $0<r<R_{in}$ (the second column in Table 1) these solutions are so-called modified Bessel functions [14].

The matching of the logarithmic derivatives of $s$-wave functions (the second line in Table 1) at the points $R_{in}$ and $R_o$ leads to the following transcendental equation for the parameters of the potential well (15), at which the $s$-level with zero binding energy $\kappa^2/2=0$ appears

$$\tan x = -\frac{\cos y + y\sin y}{\sin y - y\cos y}. \tag{19}$$

Here we introduce variables $x=qR_o$ and $y=qR_{in}$. Using wave functions in the third line of Table 1 for $p$-levels with zero energy, we obtain the following equation that connects the parameters of the potential well (15)

$$\tan x = -\frac{3j_1(y)-yj_0(y)}{3n_1(y)-yn_0(y)}. \tag{20}$$

**Table 1.** Wave functions for spherical potential of $C_{60}$ fullerene shell

| | $r<R_{in}$ | $R_{in}<r<R_o$ | $r>R_o$ |
|---|---|---|---|
| $P_{kl}(r)$ | $Akrj_l(kr)$ | $qr[Bj_l(qr)+Cn_l(qr)]$ | $kr[j_l(kr)\cos\delta_l - n_l(kr)\sin\delta_l]$ |
| $P_{\kappa 0}(r)$ | $A\sinh\kappa r$ | $qr[Bj_0(qr)+Cn_0(qr)]$ | $e^{-\kappa r}$ |
| $P_{\kappa 1}(r)$ | $A\left(\cosh\kappa r - \frac{\sinh\kappa r}{\kappa r}\right)$ | $qr[Bj_1(qr)+Cn_1(qr)]$ | $\left(1+\frac{1}{\kappa r}\right)e^{-\kappa r}$ |

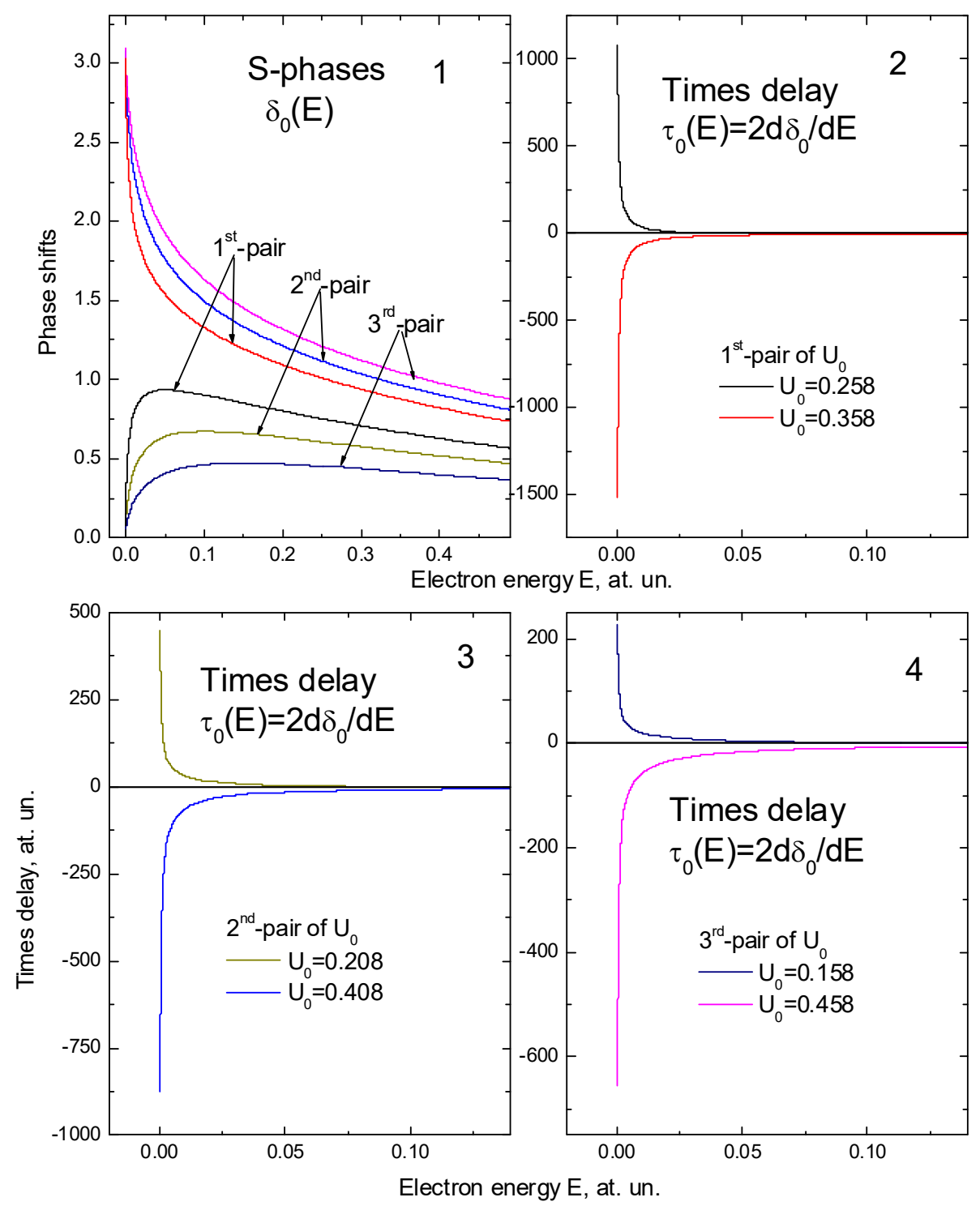

Fig.1. S-phase shifts $\delta_0(E)$ and times delay $\tau_0(E)$ (at. un.) for the different depths $U_0$ of the potential well Eq.(1)

## 4. Numerical calculations

In this section we will numerically investigate the behavior of the phases $\delta_l(E)$ and times delay as functions of $E$ in elastic electron scattering

$$\tau_l(E) = 2\frac{d\delta_l(E)}{dE}. \tag{21}$$

We will consider potential wells (1) and (15) varying the depths $U_0$ in the vicinity of the critical values $U_{0,s}$ and $U_{0,p}$ for which in these wells the first $s$- and $p$-levels $l$ appear.

### 4.1. Potential well Eq.(1)

We will assume that the radius $R_o$ of the well Eq.(1) is constant and equal to $R_o$=2. According to Eqs. (12) and (14), in such a well first $s$- and $p$- levels appear at critical depths $U_{0,s}$ =0.308 and $U_{0,p}$ =1.234, respectively.

In Fig. 1 the panel 1 presents the $s$-phase shifts for the different depths of the potential well Eq.(1). There are two families of curves in this panel. The first family corresponds to the potential well with no *s*-bound states. Three phases of this group tend to zero at $E \to 0$. These phases increase with electron energy as ~$E^{1/2}$, in accord to the Wigner threshold law. The second family of curves corresponds to the well depths $U_0$ that exceed the critical depth $U_{0,s}$=0.308. The phases of this group are equal to $\pi$ at $E$=0 and they decrease with electron energy near threshold as ~ $\pi$ -$E^{1/2}$. For the numerical calculations of the phase shifts $\delta_0(E)$ we have selected the following values of the well depth $U_0 = U_{0,s} \pm \Delta U_s$ where $\Delta U$ is equal to 0.05, 0.10 and 0.15 (that is 16, 32 and 48% of $U_s$=0.308, respectively). The lower sign minus corresponds to the first group of curves, the upper plus to the second one. In transition of the well depth $U_0$ through the critical value $U_s$ the Wigner times delay (21) experiences a jump $\Delta\tau_0(E)$. Pairwise consideration of these well depths $U_0$ allows to determine the dependence of $\Delta\tau_0(E)$ upon $\Delta U$.

The panels 2-4 in Fig. 1 are the Wigner times delay in electron elastic scattering by potential well that is calculated with the *s*-phase shifts $\delta_0(E)$ from the panel 1. Since the *s*-times delay near the threshold goes to infinity as $\tau_0(E) \propto \pm E^{-1/2}$, we will compare times delay for very small but finite value of electron energy $E$=5·10$^{-5}$ that is the starting point of our calculation. For the first pair of curves (panel 2) with $\Delta U = 0.1$ the jump of time delays (at the moment of *s*-level arising in the well) is about $\Delta\tau_0 \approx 2.6 \cdot 10^3$ (from -1.5·10$^3$ to +1.1·10$^3$). For the second pair of curves, with $\Delta U = 0.2$, panel 3, the jump of times delay is about $\Delta\tau_0 \approx 1.3 \cdot 10^3$. For the last pair of curves, that is for $\Delta U = 0.3$ (panel 4) the jump of times is about $\Delta\tau_0 \approx 8.8 \cdot 10^2$. The electron energy $E$ range where those jumps one could observe is a very narrow interval close to $E = 0$.

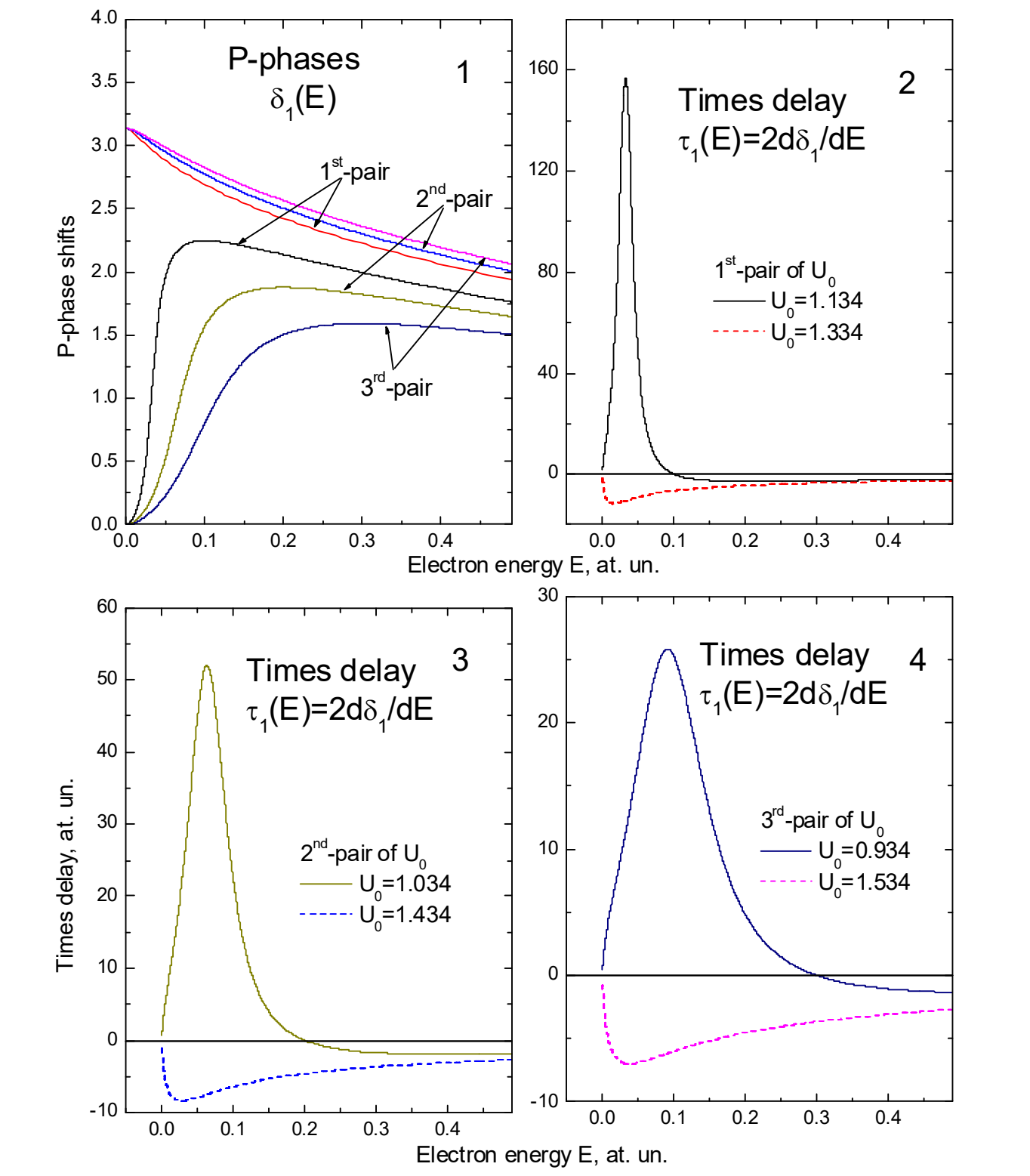

Fig. 2. P-phase shifts $\delta_1(E)$ and times delay $\tau_1(E)$ (at. un.) for the different depths $U_0$ of the potential well Eq.(1)

Panel 1 in Fig. 2 presents the *p*-phase shifts for different depths of the potential well Eq.(1). As in Fig. 2, we have here again two families of curves. The first group corresponds to the potential well with no *p*-bound states. The corresponding three lower curves go to zero at $E \to 0$. Near this point the functions $\delta_1(E)$ increase with electron energy as ~$E^{3/2}$ according to the Wigner threshold law. The second group of curves corresponds to the well depths that exceed the critical depth $U_{0,p}$=1.234. These three phases are equal to $\pi$ at the kinetic electron energy $E$=0 and they decrease with electron energy near threshold as ~ $\pi - E^{3/2}$.

For numerical calculations of the phase shifts $\delta_1(E)$ we select the following values of well depths $U_0 = U_{0,p} \pm \Delta U_p$ where $\Delta U$ is equal to 0.1, 0.2 and 0.3, that is 8,

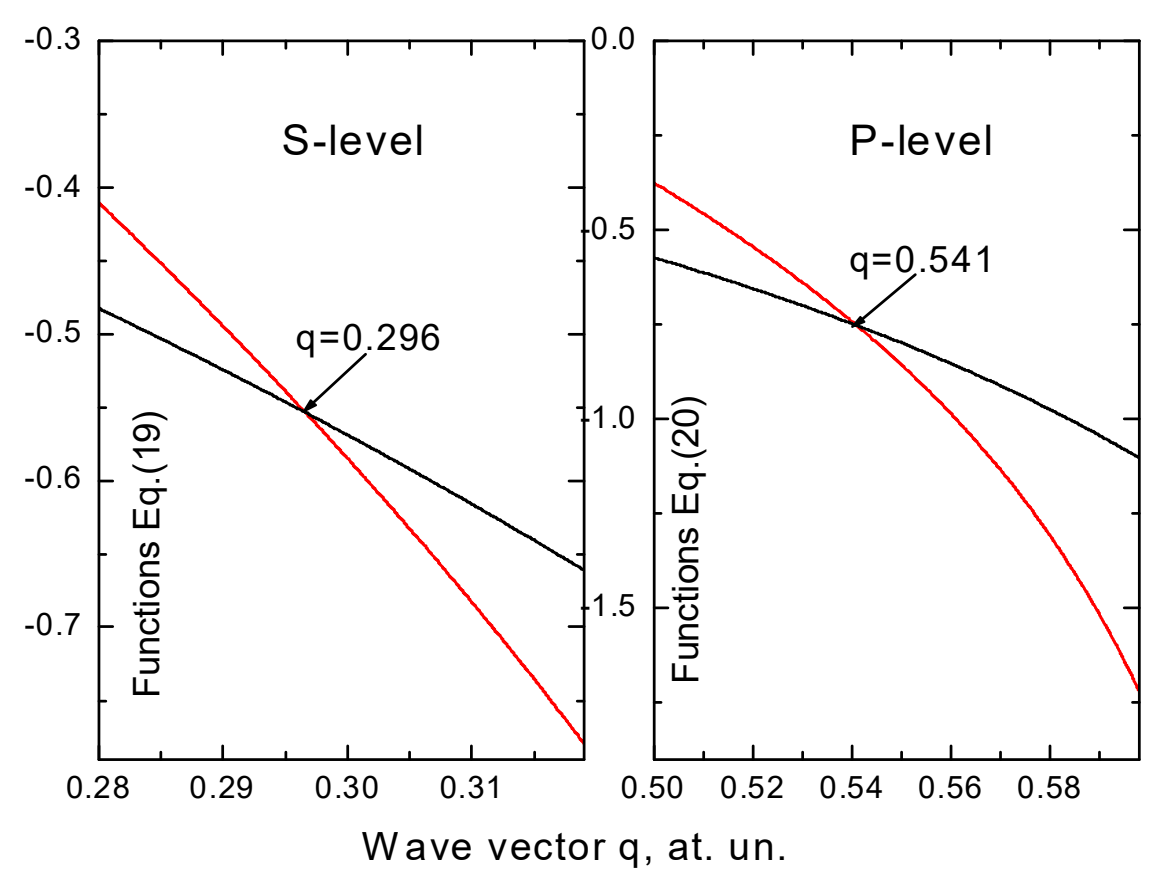

Fig. 3. The solutions of the transcendental equations (19) and (20)

16 and 24% of $U_{0,p}$=1.234. The upper sign in the expression for $U_0$ corresponds to the first group of curves; the lower plus - to the second group. In transition of the well depth $U_0$ through the critical value $U_{0,p}$ the Wigner times delay (21) experiences a jump $\Delta\tau_1(E)$.

The panels 2-4 of Fig. 2 depicts the Wigner time delays $\tau_1(E)$ that are calculated for three pairs of potential depth $U_0$ with the $p$-phase shifts $\delta_1(E)$ from the panel 1. For the first pair of curves (panel 2) with $U_0$=1.134 and 1.334 the jump of time values (at the moment of $p$-level arising in the well) is about $\Delta\tau_1 \approx 1.7\cdot 10^2$ (for electron energy $E$=0.031). Panel 3 presents the second pair of time delays, for $U_0$=1.034 and 1.434. At the solid line maximum, located at $E$=0.061 the jump of time delay (solid and dashed lines) is about ~$0.6\cdot 10^2$. For the last pair of $U_0$ ($U_0$=0.934 and 1.534) the jump of times at the maximum of solid line ($E$=0.092) is about $\Delta\tau_1 \approx 0.3\cdot 10^2$.

## 4.2. Potential well Eq.(15)

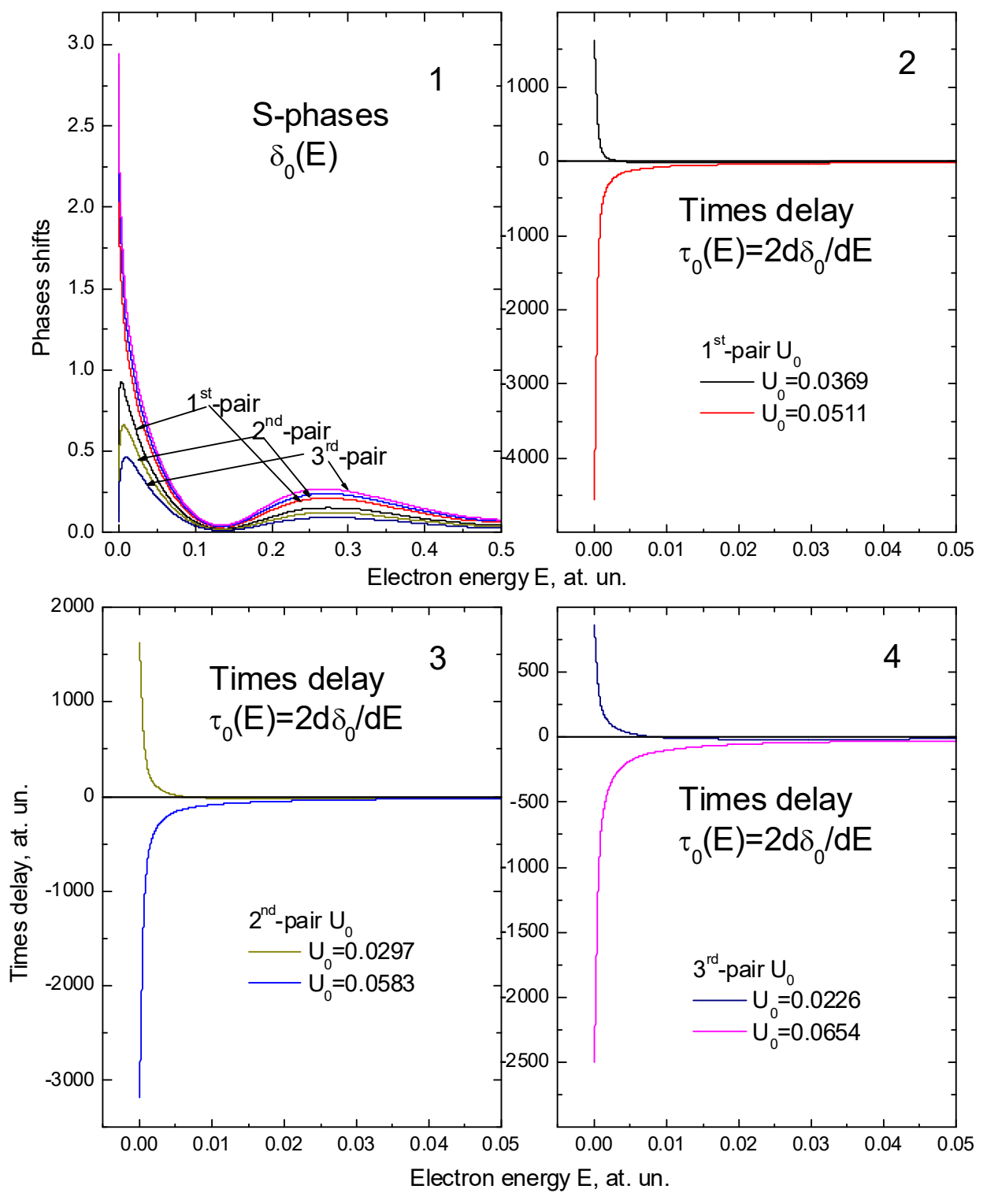

Fig. 4. S-phase shifts $\delta_0(E)$ and times delay $\tau_0(E)$ for the different depths $U_0$ of the potential well Eq.(15)

For potential well Eq. (15) we assume that the inner and outer radiuses are equal to $R_{in}$=5 and $R_o$=7, respectively. The critical values of the potential well depth that corresponds to appearing of $s$- and $p$-first levels in this well are solutions of the transcendental equations (19) and (20). The values of variable $q$ obeying Eq.(19), and thereby the critical depth of well $U_s$=$q^2$/2, are defined graphically. Fig. 3 (left panel) presents two curves corresponding to the functions being in the left and right parts of Eq.(19). The point of crossing these curves is the desired solution of Eq.(19): $q$=0.296; $U_s$=0.044. Repeating the same operations with Eq.(20) (right panel in Fig. 3), we obtain for the $p$-level the following solution: $q$=0.541; $U_p$=0.146.

Figure 4 (panel 1) presents the $s$-phase shifts for the different depths of the potential well Eq.(15). Again, we have two groups of curves. The first group corresponds to the potential well Eq.(15) with no $s$-bound states. The second group corresponds to the well depths $U_0$ that exceed the critical depth $U_{0,s}$=0.044. These potential wells have one $s$-bound state. For numerical calculations of the phase shifts $\delta_0(E)$ we employ the following values of well depths $U_0 = U_{0,s} \pm \Delta U$, where $\Delta U$ is equal to 0.0710, 0.0143 and 0.0214, that is, as in the above performed calculations, 16, 32 и 48% of $U_s$=0.044, respectively. The lower sign (minus) corresponds to the group of curves with zero values of phases at $E \to 0$ while the upper, plus, represent the second group of curves, namely that with $\delta(E)|_{E\to 0} \to \pi$.

Fig. 4 presents results similar to Fig. 1, but for the potential (15). Panels 2-4 of figure 4 are the Wigner time delays that are calculated for three pairs of potential depths $U_0$ with the $s$-phase shifts $\delta_0(E)$ from panel 1. Just as for (1), while well depths $U_0$ go through the critical values $U_{0,s}$, Wigner times delays experience jumps $\Delta\tau_0(E)$. Fig. 5 presents data, similar to Fig 4, but for $p$-phase. The corresponding phase curves in panel 1 of Fig. 5 are parallel to each other in almost all the considered energy interval. So, we can observe jumps $\Delta\tau_1(E)$ only in a very narrow range of $E \to 0$. Table 2 collects jumps in time delays at $E$=$5\cdot 10^{-5}$ for (1) and (2) potentials:

**Table 2**. Jumps of times delay $\Delta\tau_0(E)$ and $\Delta\tau_1(E)$

| | $\Delta\tau_0(E)$, at. un. Fig.1 | $\Delta\tau_0(E)$, at. un. Fig.4 | $\Delta\tau_1(E)$, at. un. Fig.2 | $\Delta\tau_1(E)$, at. un. Fig.5 |
|---|---|---|---|---|
| 1st-pair of $U_0$ | $2.6\cdot 10^3$ | $6.2\cdot 10^3$ | $1.7\cdot 10^2$ | $1.9\cdot 10^3$ |
| 2nd-pair of $U_0$ | $1.3\cdot 10^3$ | $4.8\cdot 10^3$ | $0.6\cdot 10^2$ | $7.5\cdot 10^2$ |
| 3rd-pair of $U_0$ | $0.88\cdot 10^3$ | $3.3\cdot 10^3$ | $0.3\cdot 10^2$ | $4.2\cdot 10^2$ |

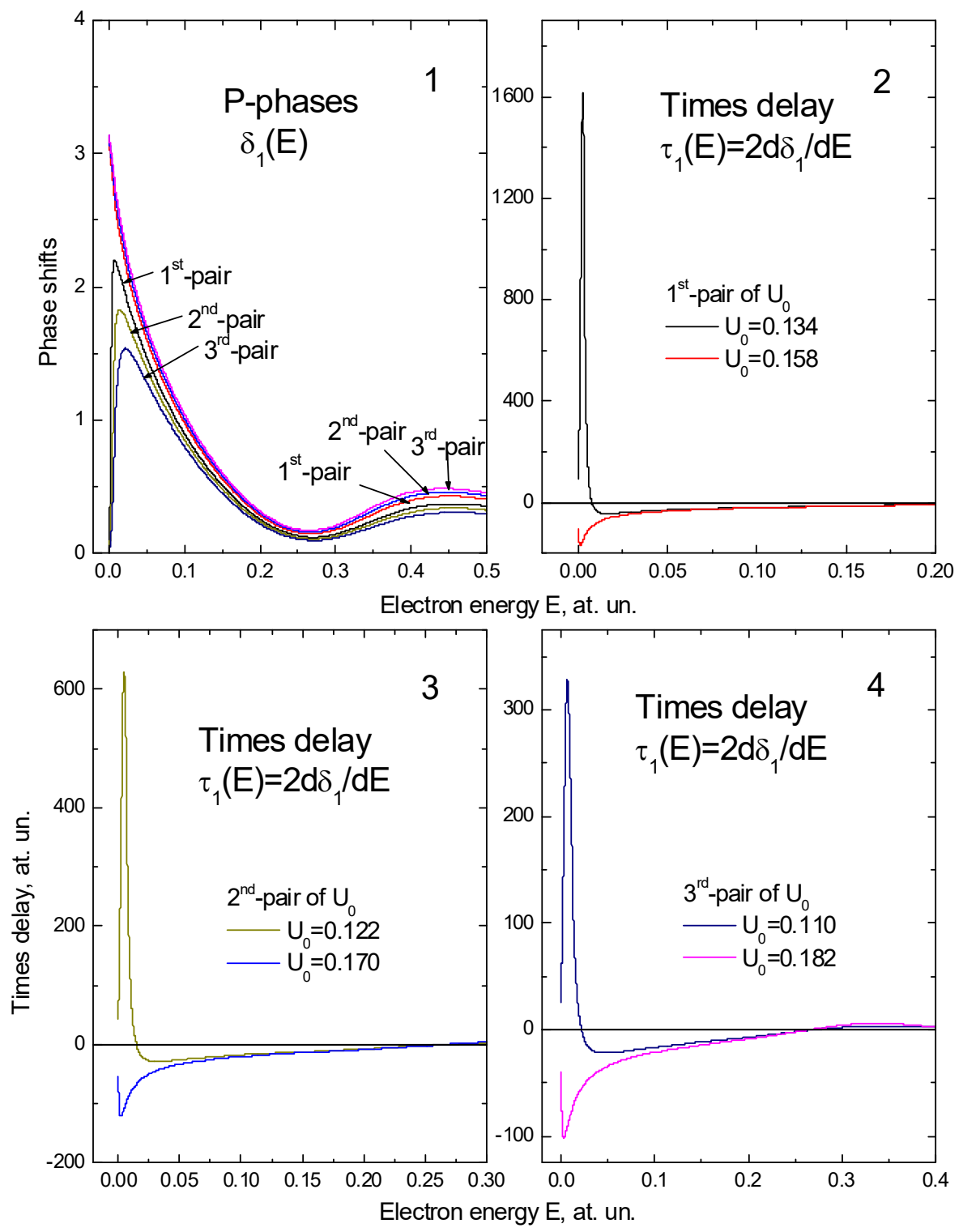

Fig. 5. P-phase shifts $\delta_1(E)$ and times delay $\tau_1(E)$ for different depths $U_0$ of the potential well Eq.(15)

The first and second columns of Table 2 collect the values of jumps for three pairs of curves for $s$-scattering, depicted in in Fig. 1 and Fig. 4, while the third and fourth column present the jumps $\Delta\tau_1(E)$ for three pairs of curves for $p$-scattering, depicted in Fig. 2 and Fig. 5, respectively. Table 2 demonstrates the systematic decrease of $\Delta\tau_l(E)$ while going from the 1st pair of $U_0$ values to the 3rd one, i.e. with growth of $\Delta U$. The data in Table 2 demonstrate that for potential (15) the jumps $\Delta\tau_1(E)$ are much bigger than for potential (1).

## 5. Conclusions and Discussions

We have investigated the phase shifts and Wigner time delay for slow electron scattering by simple, rectangular attractive potentials. We concentrated upon such parameters of these potentials that are close to their critical values, at which first $s$- and $p$ electron bound states appear. The simplicity of the considered objects permits to perform analytically an essential part of evaluations, complimenting them with some not too complex numeric calculations.

In spite of simplicity, the presented above calculations permit to draw some conclusions about special features of Wigner times delay that are universal for shallow potential wells. The time delay for potential wells with no bound levels is always positive, but changes its sign after a discrete level arises in the well. At the potential parameters at which a discrete state only arises, the Wigner times delay experiences a jump. The value of this jump $\Delta\tau_l(E)$ is increasing when deviation $\Delta U$ of the potential well depth from the critical value at which a discrete level appear, decreases.

The comparison of 2-4 panels in figures 1 and 4 (or figures 2 and 5) demonstrates that the values of times delay $\tau_l(E)$ strongly depends on geometrical sizes of potential wells. The time of wave packet transmission through the potential well (15) is several times bigger than the time delay of electron scattered by the potential well (1). To the same conclusion we came about the jumps of times delay $\Delta\tau_l(E)$. They are also much bigger for potential well (15) as compared to the well (1).

We did not touch here the problem of $\Delta\tau_l(E)$ existence and variation while approaching the second, third etc. bound levels. We expect some jumps there, but corresponding derivations are much more complex and considerably less universal.

A separate and important problem is to evaluate phase shifts and time delays in electron – atom scattering process. A lot is known about scattering phases for this process, for which relatively reliable data on ab-initio calculations exist [17]. These include also low-energy scattering. In a number of cases the addition of attractive polarization interaction alters at small $E$ the $s$-phase derivative from negative Hartree –Fock to positive values.

This alteration leads to appearance of the widely known Ramsauer minima in the cross-section of slow electron scattering by noble gas atoms. However, it is not connected there to formation of an extra bound electron level. It would be of considerable interest, and this is one of our current directions of activity, to disclose a possible connection between modifications of the atomic field and jumps in $\tau_l(E)$ on the way from a noble atom to its closest neighbor that is able to form a negative ion.

**Acknowledgments**

ASB is grateful for the support to the Uzbek Foundation Award OT-Ф2-46.

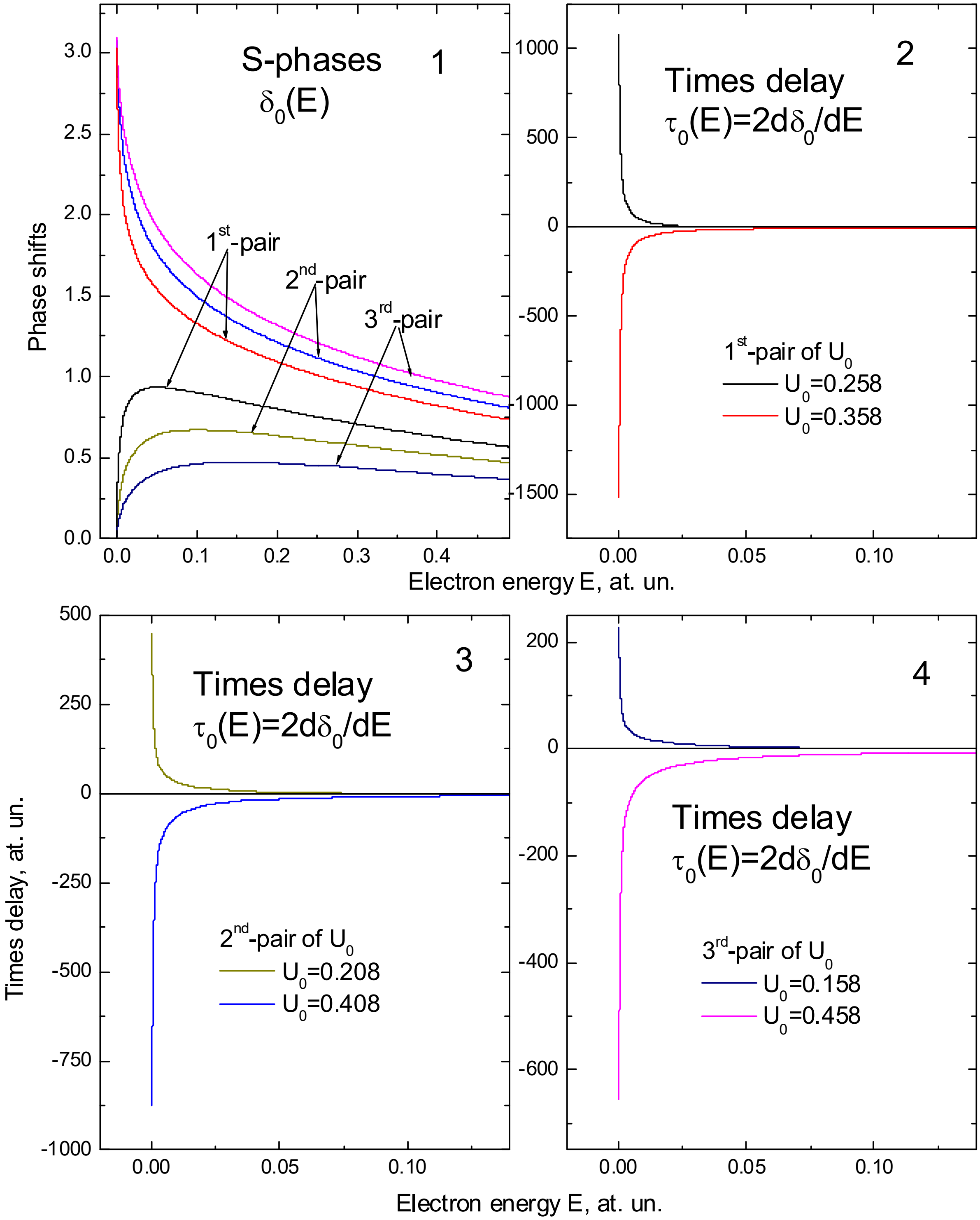

Fig.1. S-phase shifts $\delta_0(E)$ and times delay $\tau_0(E)$ (at. un.)
for the different depths $U_0$ of the potential well Eq.(1)

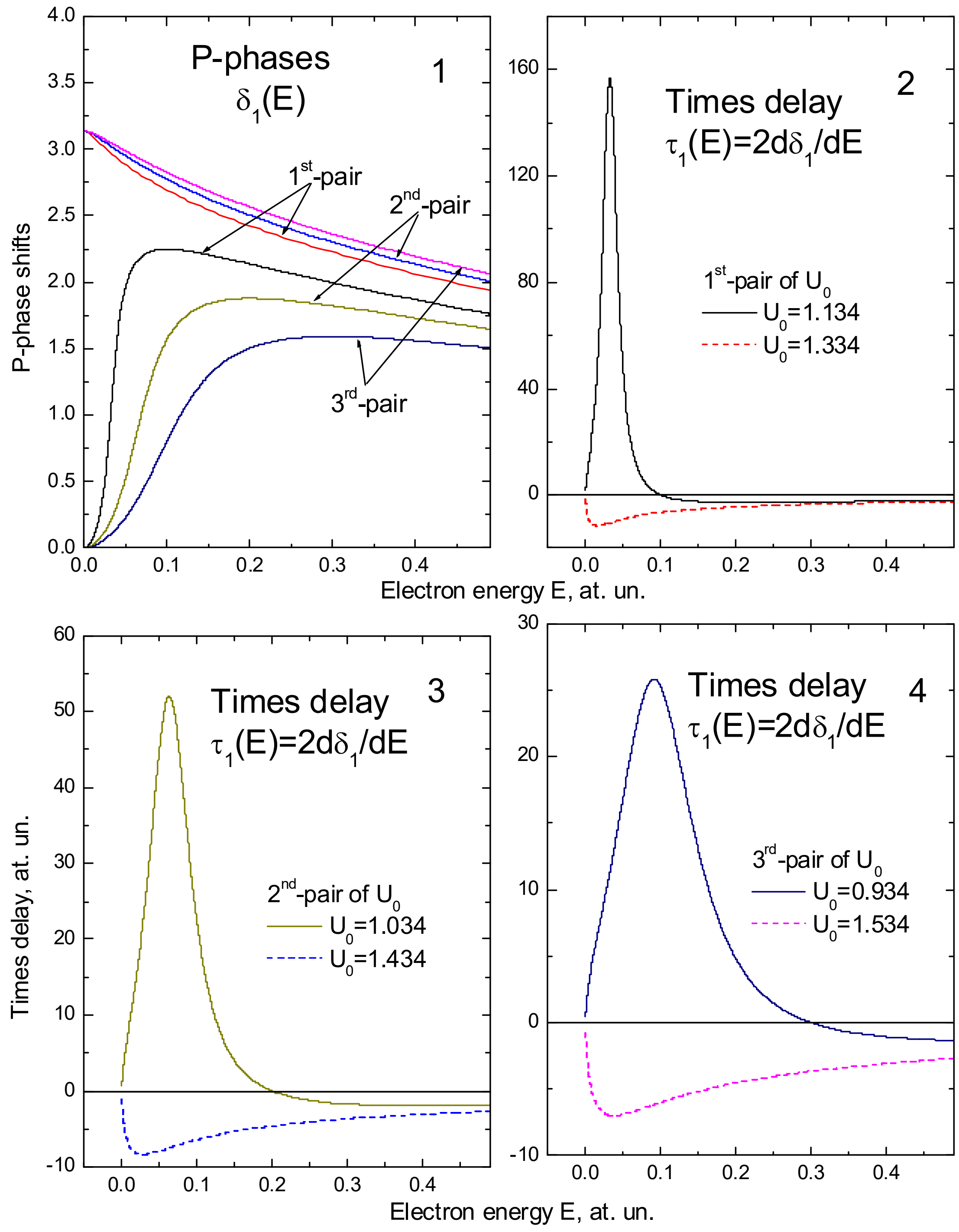

Fig. 2. P-phase shifts $\delta_1(E)$ and times delay $\tau_1(E)$ (at. un.) for the different depths $U_0$ of the potential well Eq.(1)

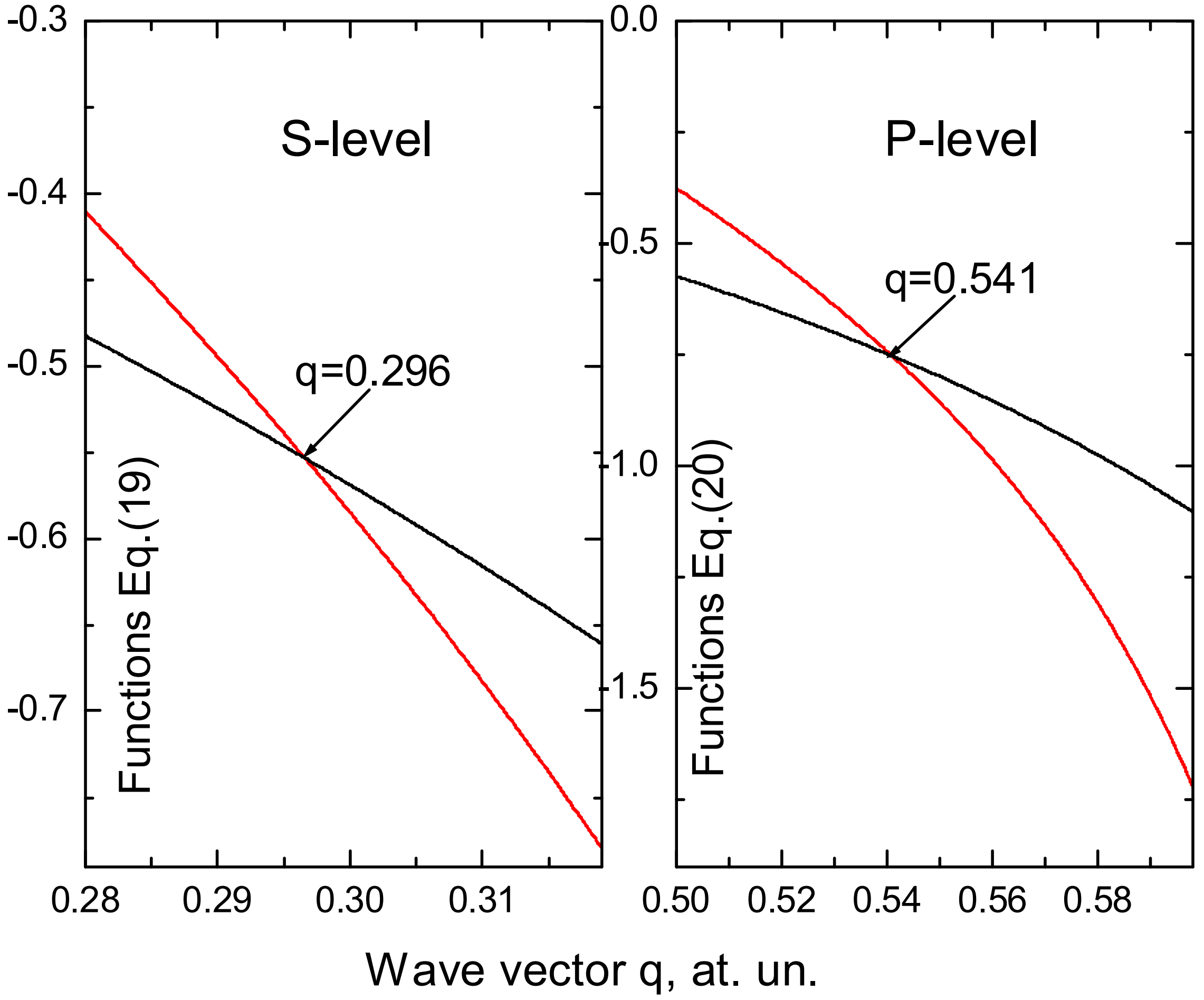

Fig. 3. The solutions of the transcendental equations (19) and (20)

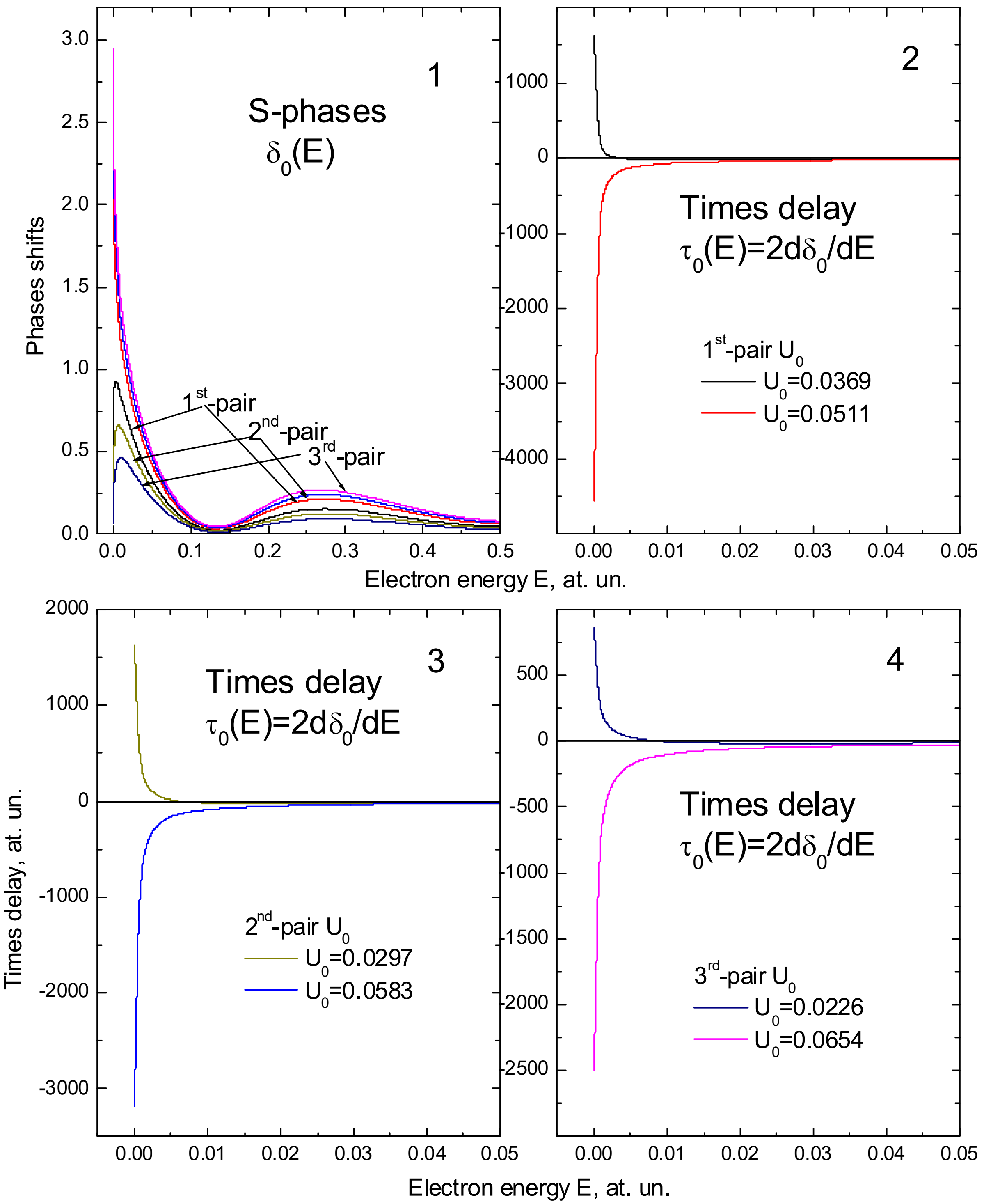

Fig. 4. S-phase shifts $\delta_0(E)$ and times delay $\tau_0(E)$ (at. un.) for the different depths $U_0$ of the potential well Eq.(15)

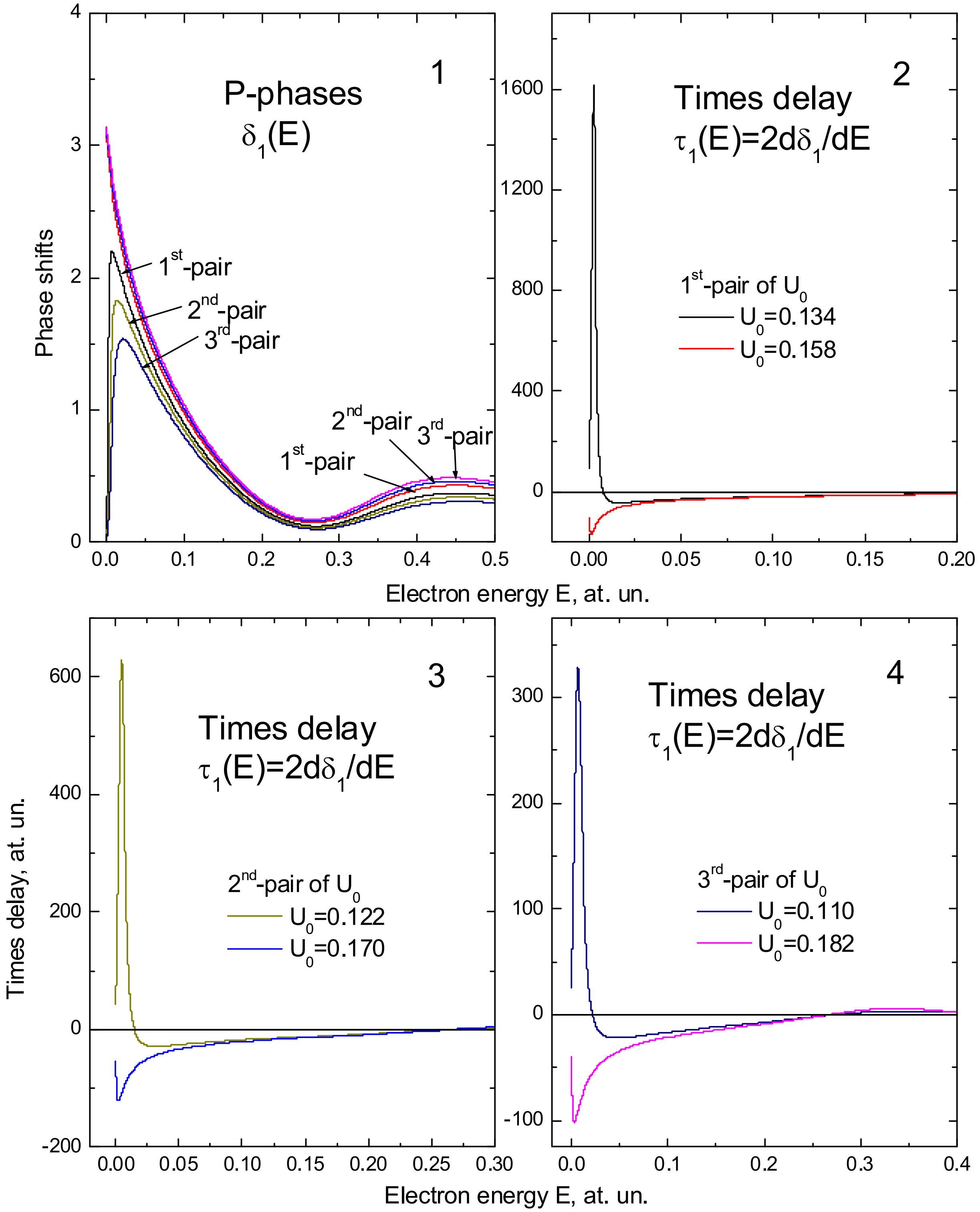

Fig. 5. P-phase shifts $\delta_1(E)$ and times delay $\tau_1(E)$ (at. un.) for the different depths $U_0$ of the potential well Eq.(15)